\documentclass[a4paper]{spie}  
\usepackage{amsmath,amsfonts,amssymb}
\usepackage{subcaption}
\usepackage[utf8]{inputenc}   
\usepackage[english]{babel}   
\usepackage[usenames,dvipsnames]{xcolor} 
\usepackage[colorlinks=true, allcolors=blue, pdftex]{hyperref}
\usepackage{aas_macros}
\usepackage{paralist}
\usepackage{graphicx}

\title{Improvements to SHINS, the SHARK-NIR Instrument Software, during the AIT phase.}

\author[*,a]{Davide~Ricci}

\author[a]{Fulvio~Laudisio}
\author[b]{Marco~De~Pascale}
\author[a]{Sona~Shivaji~Rao~Chavan}
\author[a]{Andrea~Baruffolo}

\affil[a]{INAF -- Osservatorio Astronomico di Padova, Vicolo dell’Osservatorio 5, I-35122, Padua, Italy }
\affil[b]{CINECA -- Via Magnanelli 6, I-40033, Casalecchio di Reno, Bologna, Itely}

\authorinfo{$^*$Contact information:
  D.R: davide.ricci@inaf.it, +39-049-829-3480
}

\begin{document}
\maketitle

\begin{abstract}
  
  In the context of SHARK-NIR (System for coronagraphy with High
  Order adaptive optics in Z and H band), we present the development
  of SHINS, the SHARK-NIR INstrument control Software, in particular
  focusing on the changes introduced during the Assembly, Integration,
  and Test (AIT) phase.

  SHARK-NIR observing sessions will be carried out with "ESO-style"
  Observation Blocks (OBs) based on so-called Templates scripts that
  will be prepared by observers.  We decided to develop Templates also
  for the large number of AIT tests (flexures, coronagraphic mask
  alignment, scientific camera performances...). Here we present the
  adopted HTTP API for the OBs generation and a web-based frontend
  that implements it.

  Taking advantage of this approach, we decided to expose APIs also
  for individual device movement and monitoring, as well as for
  general status. These APIs are then used in the web-based instrument
  control and synoptic panels.

  During the recent AIT phase, a potential collision issue between two
  motorized components emerged.  While we are exploring the
  possibility of a hardware interlock, we present a software solution
  developed at the Observation Software level, that is also available
  while using other software such as engineering panels. The system is
  based on three protection layers and it has been successfully
  tested.

\end{abstract}


\section{Introduction}
\label{sec:intro}

The SHARK-NIR instrument\cite{ 2016SPIE.9911E..27V,
  2016SPIE.9909E..31F, 2014SPIE.9147E..7JF}, is a forthcoming facility
for the LBT (Large Binocular Telescope), Arizona, USA.
It provides a scientific camera and a set of masks and filters mainly
for coronagraphic and direct imaging capabilities, as well as a setup
for long slit spectroscopy at low and middle
resolution\cite{2018SPIE10703E..0EF, 2018SPIE10702E..4CM,
  2018SPIE10701E..2BC}.
A fast deformable mirror in the optical path provides NCPA
corrections, residual tip-tilt correction, and the fine alignment of
the scientific target behind the coronagraphic masks, while a set of
lamps are foreseen for calibration.

SHARK-NIR successfully passed the PSR process on May 2022, and
recently approached the first run of the Pre-Commissioning phase,
which foresees the clean room re-integration at LBT.

This paper is part of a series of contributions to this conference
\cite{shark-marafatto, shark-farinato, shark-bergomi} describing the
current development status of SHARK-NIR subsystems.
In particular, we present the progress in the status of the SHARK
Instrument control Software (SHINS)\cite{2020SPIE11452E..1TD,
  2018SPIE10707E..1MD }
during the wider process of Assembly, Integration and Test
phase\cite{2020SPIE11452E..2QR, 2018SPIE10707E..1GR,
  2020SPIE11447E..5JC, 2020SPIE11448E..1MM, 2020SPIE11447E..53V,
  2020SPIE11447E..50M, 2020SPIE11447E..4RU, 2018SPIE10705E..16V}.

An overview of the software components is shown in in Sect.~\ref{sec:overview}.
The development of the generator of Observation Blocks is treated in  Sect.~\ref{sec:tobs}.
The Instrument Control panel and Synoptic panel are described in Sect.~\ref{sec:icsyn}.
The specific software system to avoid collision is presented in Sect.~\ref{sec:collision}.
Conclusions are presented in Sect.~\ref{sec:conc}.

\section{Overview}
\label{sec:overview}

SHINS instrument control software is currently composed by different
processes.  The higher level is developed in Python and it is called
sequencer (\texttt{seq}).
The sequencer interacts with the Observation Software \texttt{
  C++}-based process (\texttt{obs\_ctrl}).
The Observation Software is the central component and for such reason
it interacts will all the subsystems: the Data Manager
\texttt{C++}-based process (\texttt{data\_mgr}), and three custom
configurations of the general-purpose software \texttt{basdard}
provided by the Max Plank Institute for Astrophysics, dedicated to
motion components. Moreover, dedicated modules interfaces with the
internal tiptilt correction system, the scientific camera, and finally
the telescope.


Motion components are distinguished between: motion, calibration and
tracking, corresponding to three configurations for the control
electronic, called MoCon, developed at MPIA, that controls all the
motors. Three processes running on the sharknir workstation provide
the service to control each motorized stage and communicate with the
MoCon: the control of the motor wheels (\texttt{motion\_ctrl}), the
tracking devices (\texttt{track\_ctrl}) and the calibration unit
substystem (\texttt{calunit\_ctrl}).

A dedicated Real Time Computer called BCU (Basic Computational Unit)
controls an internal deformable mirror and a technical camera; the
service \texttt{tiptilt\_ctrl} mediates the communication between
\texttt{obs\_ctrl} and the BCU.

The scientific camera and the cryo-vacuum system are controlled by
interfacing to the specific control software provided by the Steward
Observatory written using the \texttt{INDI} protocol, house-developed
at LBT, running on a separate workstation.  The scientific camera
workstation also hosts a replica of the LBTI instrument webpage as a
helper tool during commissioning. It will be used only to control AO
and to send telescope presets. It could become a definitive
configuration depending on the outcome of commissioning time using
this tool.

Finally, the service \texttt{tcs\_ctrl} mediates the communication
with the telescope.

\section{Templates and Observation Blocks}
\label{sec:tobs}

\begin{figure}[t]
  \centering
  \begin{minipage}{.495\linewidth}
    \begin{subfigure}[t]{.95\linewidth}
      \includegraphics[width=\textwidth]{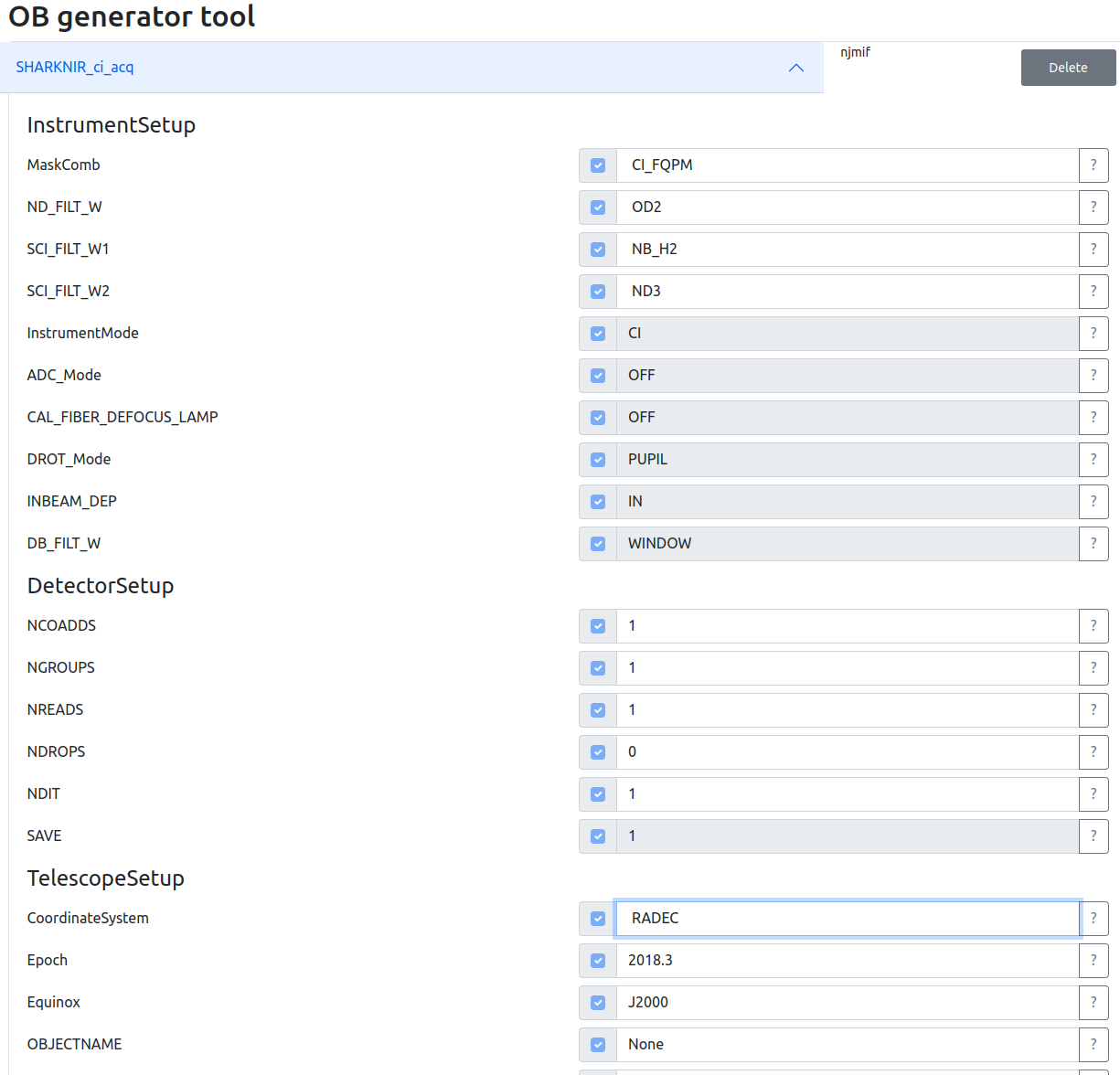}
      \caption{Instance of the \texttt{ci\_acq} template in a new OB.}
      \label{fig:tobs-a}
    \end{subfigure}
  \end{minipage}
  \begin{minipage}{.495\linewidth}
    \begin{subfigure}[t]{.95\linewidth}
      \includegraphics[width=\textwidth]{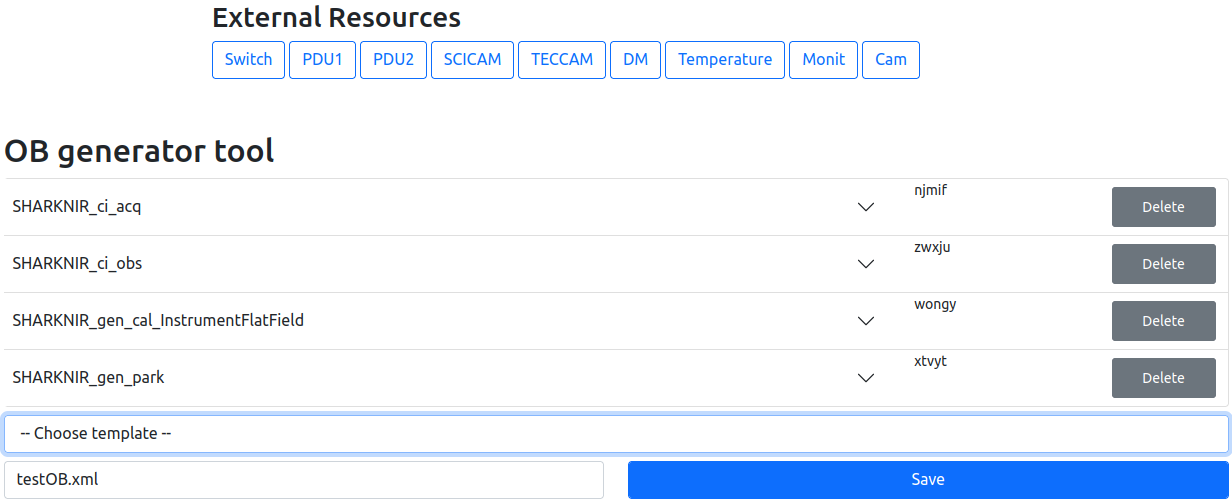}
      \caption{OB containing 4 template instances.}
      \label{fig:tobs-b}
    \end{subfigure} \\
    \begin{subfigure}[b]{.95\linewidth}
      \includegraphics[width=\textwidth]{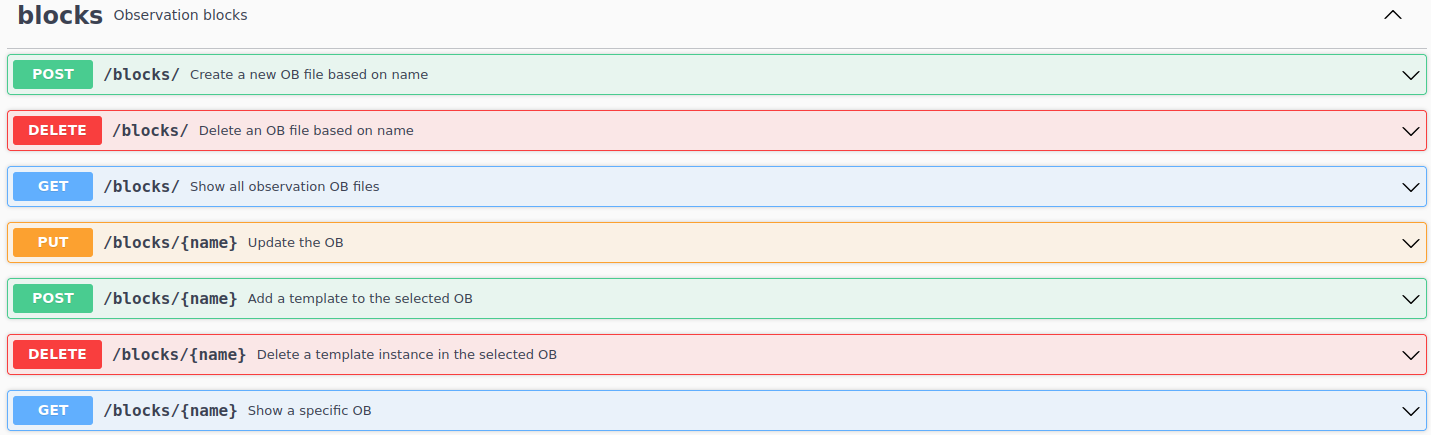}
      \caption{Swagger documentation of the API calls.}
      \label{fig:tobs-c}
    \end{subfigure} 
  \end{minipage}
  \caption{Screenshot of the OB generator tool and API documentation.}
  \label{fig:tobs}
\end{figure}

SHARK-NIR is thought to be operated by means of Observing Blocks
(OBs), in the same way this is done with instruments hosted in modern
astronomical observatories such as the Very Large Telescope (VLT).
  
The underlying communication engine between the processes is Internet
Communication Engine (ICE) developed by company ZeroC. A
service-oriented architecture framework, called Twice As Nice
(\texttt{TaN}), developed at MPIA-Heidelberg, mediates the
communication with the \texttt{basdard} processes.

OBs are stored as XML files; the content specifies the sequence of
scripts for automated operations, called Templates, and the parameters
to be passed to them. Each template implements an observation,
calibration or maintenance procedure. It is composed by a reference
file, describing all the parameters which are fixed during the
execution of the template; a signature file, describing all the
parameters the user can specify; and a template script, which is the
actual implementation of the procedure.
OBs can be prepared with any XML generator software, for example the
Observation Tool (OT) generator of the LBT Observatory.

Templates scripts are implemented as Python scripts accepting four
input arguments in the form of Python dictionary, an unordered list of
labels (representing the parameters) and values. Each dictionary
specifies the setups as follows: Instrument setup; Detector setup;
Telescope setup; Real-time tip/tilt subsystem setup.
When an OB is loaded, \texttt{ seq} process operates as follows: reads
the specified XML files and extract an ordered list of templates to
execute; for each template, the corresponding list of parameters is
then checked against its scheme. The resulting Python dictionaries are
merged with the ones containing the fixed parameters to obtain a
complete setup.  Finally, the OB is executed.  During execution,
\texttt{seq} calls the list of Python scripts in the order specified
by the OB, passing the corresponding dictionary objects as parameter.

In order to test observation, calibration and maintenance templates
before the commissioning at LBT, a web-based generator of OB in XML
format was developed. The webtool is shown in \ref{fig:tobs}. It is
based on a Representational State Transfer Application Programming
Interface (REST API) that use the common HTTP verbs (\texttt{GET} to
retrieve a resource, \texttt{PUT} to modify it, \texttt{POST} to
create a new resource, \texttt{DELETE} to delete it.) The Python
package \texttt{Flask RestX} was used to build the API, in order to
provide automatic documentation generated by the \texttt{Swagger}
module\footnote{\url{https://swagger.io/}}, a set of tools that
simplify API development and documentation for users, teams, and
enterprises.

On top of the API, a web interface was built. The interface allows:
\begin{inparaenum}
\item the creation of a new OB;
\item the addition of templates within an OB; 
\item a quick editing of the free template parameters, as well as a
  check of the value of the fixed parameters;
\item drag and drop reordering of the templates;
\item the possibility to choose a name for OB saving.
\end{inparaenum}

\section{Instrument Control panel and Synoptic panel}
\label{sec:icsyn}

\begin{figure} [t]
  \centering
  \begin{tabular}{cc}
    \includegraphics[width=0.99\textwidth]{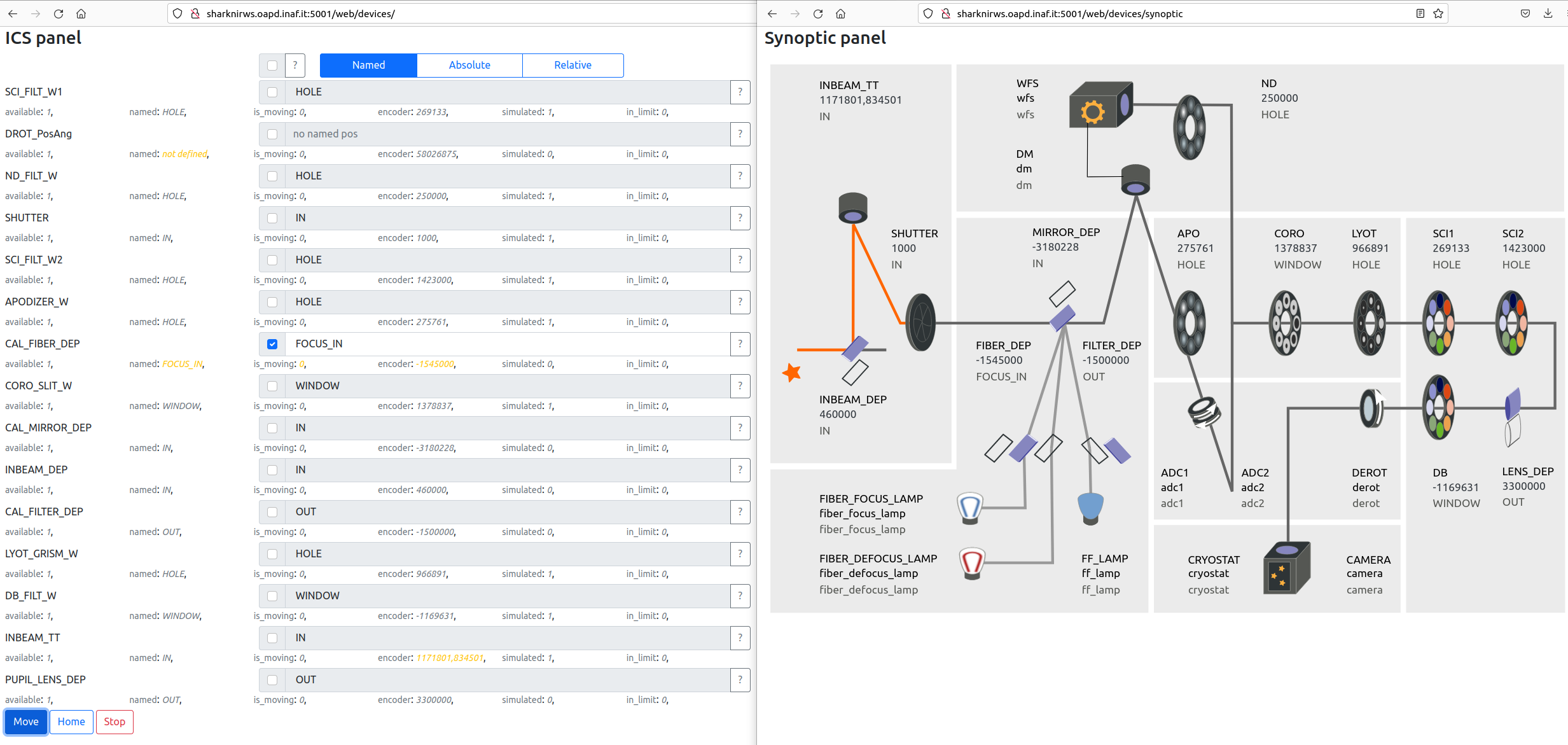}
  \end{tabular}
  \caption[AFC]
  { \label{fig:icsyn}
    Left: Instrument Control panel.
    Right: Synoptic panel.}
\end{figure}

SHARK-NIR uses the \texttt{TaN} framework to manage all motor devices,
which are divided in \texttt{calunit\_ctrl}, \texttt{motion\_ctrl},
and \texttt{track\_ctrl} services. This framework provides a low-level
Graphical User Interface (GUI) developed in Qt, for each one of these
services. Each one of these GUIs allows to select one device at the
time, to move it in the home or named positions, or perform absolute
or relative movements.

While this solution is robust and acts as a low level software
interface (i.e. at a lower level with respect the OS software layer),
it does not allow a global overview of the devices.

For this reason, we decided to develop a ``global'' Instrument Control
panel for SHARK-NIR.  We chose to go on with the APIs approach already
used for the OB generator tool, exposing HTTP methods to get and
manage the status of devices. This means that these controls pass
through the OS layer, as opposite to what happens with \texttt{TaN}
GUIs.

On top of these APIs we developed a web-based visualization tool also
presented in this conference\cite{oarpaf-ricci} (see
Fig.~\ref{fig:icsyn}, Left). The API methods allow enough modularity
to show ad a simple web form all or individual devices, so that it is
possible to re-use the visualization of a singular ``web form device''
as snippet in other visualizations, or use it ``detached'' as a popup,
if needed.

The Synoptic panel (see Fig.~\ref{fig:icsyn}, Right) is used to show a
graphical representation of the instrument's hardware devices, in
order for the operator to have a quick view of the setup. It does not
reflects the real spatial displacement of the components; it shows the
devices as they are crossed by the optical path and makes this path
easier to understand.

Moreover, the synoptic panel provides a base level of animation for
some devices; for instance: the shutter image can be open or closed,
and the deployers' position changes on the basis of the setup. Text
fields can show the motors position (expressed in steps) and their
named position, if applicable.

Finally, the optical path changes its color based on the setup, to
enhance whether the light source is coming from a specific calibration
lamp or from a source on sky. In Fig.~\ref{fig:icsyn} the beam
deployer is inserted and the shutter is closed, so the light path
stops at the shutter level.

\section{Anti-collision software}
\label{sec:collision}

During the recent AIT phase, a potential collision issue between two
motorized components emerged, namely the coronagraphic motor wheel and
the fiber deployer.  While we are exploring the possibility of a
hardware interlock, we developed a software solution at the
Observation Software level, that is also available while using other
software such as engineering panels. The anti-collision system is
based on three protection layers:

\begin{enumerate}
  
\item The first level of protection consists in the fact that the
  wheel before making any movement checks the position of the fiber
  deployer, and if this is below a certain threshold it raises an
  exception.

\item The second level of protection consists of a couple of functions
  that intervene during the setup phase of the instrument. The setup
  is classified according to wheel and deployer movements, and is
  divided into nine scenarios. Of these nine scenarios, five don't
  need any special attention, while for the last four the software
  splits the setup into two or three consecutive steps where the
  coronagraphic wheel never moves when the fiber is in the cut-off
  region, minimizing execution times. Eventually, the fiber is moved
  to the safe region and then returned to the initial position at the
  end of the wheel setup.

\item The third level of protection is based on a set of callbacks
  installed on the services controlling the motors. Whenever the
  coronagraphic slit wheel starts any movement the framework calls
  back the Observation Software which in turn launchs a new thread
  that controls continously the calibration fiber deployer
  position. If it is beyond the interdiction threshold, it commands an
  abort of the movement and moves the fiber deployer into the safe
  area. This third level of protection intervenes whenever a movement
  of the coronagraphic wheel is commanded by \textit{any} client of
  the motor services, provided that \textit{SHINS} is up and running,
  thus protecting the instrument also from accidental errors of an
  operator working with the engineering guis.
  
\end{enumerate}

The anti-collision system is driven by the Observation Software, and
relies on the position of the calibration fiber deployer; for this
reason at the startup of the Observation Software such deployer is
always homed.

\section{Conclusion}
\label{sec:conc}

We presented the progresses in the development of the Instrument
Control Software of the forthcoming SHARK-NIR instrument.
During the AIT phase, several unattended necessities emerged.

The first one was to make up for the lack of an OB generator tool.
We then developed a web interface built over HTTP APIs to generate and
edit Observation blocks based on Templates.

The second was relative of the risk of collision between two motorized
components. Being these components driven by two separate processes,
an anti-collision software was developed ad the Observation Software
level to manage all possible scenarios.

The effort of developing solutions for these problems allowed us to
develop solutions flexible enough to be also used for the
implementation of a web-based Instrument Control panel and a Synoptic
panel, that will be used during all the commissioning phase, and
surely during the subsequent operations at the telescope.


\bibliographystyle{spiebib} 
\bibliography{ricci-shark} 

\begin{thebibliography}{10}

\bibitem{2016SPIE.9911E..27V}
{Viotto}, V., {Farinato}, J., {Greggio}, D., {Vassallo}, D., {Carolo}, E.,
  {Baruffolo}, A., {Bergomi}, M., {Carlotti}, A., {De Pascale}, M., {D'Orazi},
  V., {Fantinel}, D., {Magrin}, D., {Marafatto}, L., {Mohr}, L., {Ragazzoni},
  R., {Salasnich}, B., and {Verinaud}, C., ``{SHARK-NIR system design analysis
  overview},'' in [{\em Modeling, Systems Engineering, and Project Management
  for Astronomy VI}{\nolinebreak\hspace{0.1em}]},  {Angeli}, G.~Z. and
  {Dierickx}, P., eds., {\em Society of Photo-Optical Instrumentation Engineers
  (SPIE) Conference Series} {\bf 9911},  991127 (Aug. 2016).

\bibitem{2016SPIE.9909E..31F}
{Farinato}, J., {Bacciotti}, F., {Baffa}, C., {Baruffolo}, A., {Bergomi}, M.,
  {Bongiorno}, A., {Carbonaro}, L., {Carolo}, E., {Carlotti}, A., {Centrone},
  M., {Close}, L., {De Pascale}, M., {Dima}, M., {D'Orazi}, V., {Esposito}, S.,
  {Fantinel}, D., {Farisato}, G., {Gaessler}, W., {Giallongo}, E., {Greggio},
  D., {Guyon}, O., {Hinz}, P., {Lisi}, F., {Magrin}, D., {Marafatto}, L.,
  {Mohr}, L., {Montoya}, M., {Pedichini}, F., {Pinna}, E., {Puglisi}, A.,
  {Ragazzoni}, R., {Salasnich}, B., {Stangalini}, M., {Vassallo}, D.,
  {Verinaud}, C., and {Viotto}, V., ``{SHARK-NIR: from K-band to a key
  instrument, a status update},'' in [{\em Adaptive Optics Systems
  V}{\nolinebreak\hspace{0.1em}]},  {Marchetti}, E., {Close}, L.~M., and
  {V{\'e}ran}, J.-P., eds., {\em Society of Photo-Optical Instrumentation
  Engineers (SPIE) Conference Series} {\bf 9909},  990931 (July 2016).

\bibitem{2014SPIE.9147E..7JF}
{Farinato}, J., {Pedichini}, F., {Pinna}, E., {Baciotti}, F., {Baffa}, C.,
  {Baruffolo}, A., {Bergomi}, M., {Bruno}, P., {Cappellaro}, E., {Carbonaro},
  L., {Carlotti}, A., {Centrone}, M., {Close}, L., {Codona}, J., {Desidera},
  S., {Dima}, M., {Esposito}, S., {Fantinel}, D., {Farisato}, G., {Fontana},
  A., {Gaessler}, W., {Giallongo}, E., {Gratton}, R., {Greggio}, D., {Guerra},
  J.~C., {Guyon}, O., {Hinz}, P., {Leone}, F., {Lisi}, F., {Magrin}, D.,
  {Marafatto}, L., {Munari}, M., {Pagano}, I., {Puglisi}, A., {Ragazzoni}, R.,
  {Salasnich}, B., {Sani}, E., {Scuderi}, S., {Stangalini}, M., {Testa}, V.,
  {Verinaud}, C., and {Viotto}, V., ``{SHARK (System for coronagraphy with High
  order Adaptive optics from R to K band): a proposal for the LBT 2nd
  generation instrumentation},'' in [{\em Ground-based and Airborne
  Instrumentation for Astronomy V}{\nolinebreak\hspace{0.1em}]},  {Ramsay},
  S.~K., {McLean}, I.~S., and {Takami}, H., eds., {\em Society of Photo-Optical
  Instrumentation Engineers (SPIE) Conference Series} {\bf 9147},  91477J (Aug.
  2014).

\bibitem{2018SPIE10703E..0EF}
{Farinato}, J., {Agapito}, G., {Bacciotti}, F., {Baffa}, C., {Baruffolo}, A.,
  {Bergomi}, M., {Bianco}, A., {Bongiorno}, A., {Carbonaro}, L., {Carolo}, E.,
  {Carlotti}, A., {Chinellato}, S., {Close}, L., {De Pascale}, M., {Dima}, M.,
  {D'Orazi}, V., {Esposito}, S., {Fantinel}, D., {Farisato}, G., {Gaessler},
  W., {Giallongo}, E., {Greggio}, D., {Guyon}, O., {Hinz}, P., {Lessio}, L.,
  {Magrin}, D., {Marafatto}, L., {Mesa}, D., {Mohr}, L., {Montoya}, M.,
  {Pedichini}, F., {Pinna}, E., {Puglisi}, A., {Ragazzoni}, R., {Salasnich},
  B., {Stangalini}, M., {Vassallo}, D., {V{\'e}rinaud}, C., {Viotto}, V., and
  {Zanutta}, A., ``{SHARK-NIR: the coronagraphic camera for LBT in the AIV
  phase at INAF-Padova},'' in [{\em Adaptive Optics Systems
  VI}{\nolinebreak\hspace{0.1em}]},  {Close}, L.~M., {Schreiber}, L., and
  {Schmidt}, D., eds., {\em Society of Photo-Optical Instrumentation Engineers
  (SPIE) Conference Series} {\bf 10703},  107030E (July 2018).

\bibitem{2018SPIE10702E..4CM}
{Marafatto}, L., {Bergomi}, M., {Biondi}, F., {Carolo}, E., {Chinellato}, S.,
  {De Pascale}, M., {Dima}, M., {Farinato}, J., {Greggio}, D., {Lessio}, L.,
  {Magrin}, D., {Portaluri}, E., {Ragazzoni}, R., {Vassallo}, D., and {Viotto},
  V., ``{The AIV concept of SHARK-NIR, a new coronagraph for the Large
  Binocular Telescope},'' in [{\em Ground-based and Airborne Instrumentation
  for Astronomy VII}{\nolinebreak\hspace{0.1em}]},  {Evans}, C.~J., {Simard},
  L., and {Takami}, H., eds., {\em Society of Photo-Optical Instrumentation
  Engineers (SPIE) Conference Series} {\bf 10702},  107024C (July 2018).

\bibitem{2018SPIE10701E..2BC}
{Carolo}, E., {Vassallo}, D., {Farinato}, J., {Agapito}, G., {Bergomi}, M.,
  {Biondi}, F., {Chinellato}, S., {Carlotti}, A., {De Pascale}, M., {Dima}, M.,
  {D'Orazi}, V., {Greggio}, D., {Magrin}, D., {Marafatto}, L., {Mesa}, D.,
  {Pinna}, E., {Portaluri}, E., {Puglisi}, A., {Ragazzoni}, R., {Stangalini},
  M., {Umbriaco}, G., and {Viotto}, V., ``{SHARK-NIR coronagraphic simulations:
  performance dependence on the Strehl ratio},'' in [{\em Optical and Infrared
  Interferometry and Imaging VI}{\nolinebreak\hspace{0.1em}]},
  {Creech-Eakman}, M.~J., {Tuthill}, P.~G., and {M{\'e}rand}, A., eds., {\em
  Society of Photo-Optical Instrumentation Engineers (SPIE) Conference Series}
  {\bf 10701},  107012B (July 2018).

\bibitem{shark-marafatto}
{Marafatto}, L. et~al., ``{SHARK-NIR on its way to LBT},'' in [{\em
  {Ground-based and Airborne Instrumentation for Astronomy
  IX}}{\nolinebreak\hspace{0.1em}]},  {\em \procspie} {\bf 12184-142} (2022).

\bibitem{shark-farinato}
{Farinato}, J. et~al., ``{Adaptive Optics Systems VIII},'' in [{\em {Software
  and Cyberinfrastructure for Astronomy VII }}{\nolinebreak\hspace{0.1em}]},
  {\em \procspie} {\bf 12185-74} (2022).

\bibitem{shark-bergomi}
{Bergomi}, M. et~al., ``{Modeling, Systems Engineering, and Project Management
  for Astronomy X},'' in [{\em {Software and Cyberinfrastructure for Astronomy
  VII }}{\nolinebreak\hspace{0.1em}]},  {\em \procspie} {\bf 12187-8} (2022).

\bibitem{2020SPIE11452E..1TD}
{De Pascale}, M., {Baruffolo}, A., {Salasnich}, B., {Ricci}, D., {Briegel}, F.,
  {Farinato}, J., {Biondi}, F., {Grenz}, P., and {Vassallo}, D., ``{SHARK-NIR:
  implementation of the instrument control software SHINS},'' in [{\em Society
  of Photo-Optical Instrumentation Engineers (SPIE) Conference
  Series}{\nolinebreak\hspace{0.1em}]},  {\em Society of Photo-Optical
  Instrumentation Engineers (SPIE) Conference Series} {\bf 11452},  114521T
  (Dec. 2020).

\bibitem{2018SPIE10707E..1MD}
{De Pascale}, M., {Baruffolo}, A., {Salasnich}, B., {Bergomi}, M., {Briegel},
  F., {D'Orazi}, V., {Downey}, E.~C., {Farinato}, J., {Hinz}, P.~M.,
  {Marafatto}, L., {Mohr}, L., and {Viotto}, V., ``{Design of SHINS: the
  SHARK-NIR instrument control software},'' in [{\em Software and
  Cyberinfrastructure for Astronomy V}{\nolinebreak\hspace{0.1em}]},  {Guzman},
  J.~C. and {Ibsen}, J., eds., {\em Society of Photo-Optical Instrumentation
  Engineers (SPIE) Conference Series} {\bf 10707},  107071M (July 2018).

\bibitem{2020SPIE11452E..2QR}
{Ricci}, D., {Baruffolo}, A., {Salasnich}, B., {De Pascale}, M., {Campana}, S.,
  {Claudi}, R., {Schipani}, P., {Aliverti}, M., {Ben-Ami}, S., {Biondi}, F.,
  {Capasso}, G., {Cosentino}, R., {D'Alessio}, F., {D'Avanzo}, P., {Hershko},
  O., {Kuncarayakti}, H., {Landoni}, M., {Munari}, M., {Pignata}, G., {Rubin},
  A., {Scuderi}, S., {Vitali}, F., {Young}, D., {Achr{\'e}n}, J.~M.,
  {Araiza-Dur{\'a}n}, J.~A., {Arcavi}, I., {Brucalassi}, A., {Bruch}, R.,
  {Cappellaro}, E., {Colapietro}, M., {Della Valle}, M., {Di Benedetto}, R.,
  {D'Orsi}, S., {Gal-Yam}, A., {Genoni}, M., {Hernandez}, M., {Kotilainen}, J.,
  {Li Causi}, G., {Mattila}, S., {Rappaport}, M., {Riva}, M., {Smartt}, S.,
  {Sanchez}, R.~Z., {Stritzinger}, M., {Ventura}, H., and {Radhakrishnan}, K.,
  ``{Development status of the SOXS instrument control software},'' in [{\em
  Society of Photo-Optical Instrumentation Engineers (SPIE) Conference
  Series}{\nolinebreak\hspace{0.1em}]},  {\em Society of Photo-Optical
  Instrumentation Engineers (SPIE) Conference Series} {\bf 11452},  114522Q
  (Dec. 2020).

\bibitem{2018SPIE10707E..1GR}
{Ricci}, D., {Baruffolo}, A., {Salasnich}, B., {Fantinel}, D., {Urrutia}, J.,
  {Campana}, S., {Claudi}, R., {Schipani}, P., {Aliverti}, M., {Ben-Ami}, S.,
  {Biondi}, F., {Brucalassi}, A., {Capasso}, G., {Cosentino}, R., {D'Alessio},
  F., {D'Avanzo}, P., {Diner}, O., {Kuncarayakti}, H., {Munari}, M., {Rubin},
  A., {Scuderi}, S., {Vitali}, F., {Achr{\'e}n}, J., {Araiza-Dur{\'a}n}, J.~A.,
  {Arcavi}, I., {Bianco}, A., {Cappellaro}, E., {Colapietro}, M., {Della
  Valle}, M., {D'Orsi}, S., {Fynbo}, J., {Gal-Yam}, A., {Genoni}, M.,
  {Hirvonen}, M., {Kotilainen}, J., {Kumar}, T., {Landoni}, M., {Lehti}, J.,
  {Li Causi}, G., {Marafatto}, L., {Mattila}, S., {Pariani}, G., {Pignata}, G.,
  {Rappaport}, M., {Riva}, M., {Smartt}, S., {Turatto}, M., and {Z{\'a}nmar
  S{\'a}nchez}, R., ``{Architecture of the SOXS instrument control software},''
  in [{\em Software and Cyberinfrastructure for Astronomy
  V}{\nolinebreak\hspace{0.1em}]},  {Guzman}, J.~C. and {Ibsen}, J., eds., {\em
  Society of Photo-Optical Instrumentation Engineers (SPIE) Conference Series}
  {\bf 10707},  107071G (July 2018).

\bibitem{2020SPIE11447E..5JC}
{Cabona}, L., {Ricci}, D., {Marini}, A., {Santostefano}, M., {Aliverti}, M.,
  {La Camera}, A., {Righi}, C., and {Tosi}, S., ``{Cerberus: A three-headed
  instrument for the OARPAF telescope},'' in [{\em Society of Photo-Optical
  Instrumentation Engineers (SPIE) Conference
  Series}{\nolinebreak\hspace{0.1em}]},  {\em Society of Photo-Optical
  Instrumentation Engineers (SPIE) Conference Series} {\bf 11447},  114475J
  (Dec. 2020).

\bibitem{2020SPIE11448E..1MM}
{Marafatto}, L., {Bergomi}, M., {Biondi}, F., {Carolo}, E., {De Pascale}, M.,
  {Greggio}, D., {Lessio}, L., {Mesa}, D., {Radhakrishnan Santhakumari}, K.~K.,
  {Umbriaco}, G., {Vassallo}, D., {Viotto}, V., {Bianco}, A., {Dima}, M.,
  {D'Orazi}, V., {Grenz}, P., {Leisenring}, J.~M., {Mohr}, L., {Montoya}, M.,
  {Zanutta}, A., {Antoniucci}, S., {Arcidiacono}, C., {Bacciotti}, F., {Baffa},
  C., {Baruffolo}, A., {Bongiorno}, A., {Carlotti}, A., {Chinellato}, S.,
  {Close}, L., {Di Filippo}, S., {Esposito}, S., {Farisato}, G., {Guyon}, O.,
  {Hinz}, P., {Magrin}, D., {Pedichini}, F., {Pinna}, E., {Portaluri}, E.,
  {Puglisi}, A., {Ragazzoni}, R., {Rossi}, F., and {Farinato}, J.,
  ``{SHARK-NIR, toward the installation at the Large Binocular Telescope},'' in
  [{\em Society of Photo-Optical Instrumentation Engineers (SPIE) Conference
  Series}{\nolinebreak\hspace{0.1em}]},  {\em Society of Photo-Optical
  Instrumentation Engineers (SPIE) Conference Series} {\bf 11448},  114481M
  (Dec. 2020).

\bibitem{2020SPIE11447E..53V}
{Vassallo}, D., {Bergomi}, M., {Biondi}, F., {Carolo}, E., {Greggio}, D.,
  {Marafatto}, L., {Umbriaco}, G., {Baruffolo}, A., {De Pascale}, M., {Plenz},
  M., {Radhakrishnan}, K., {Viotto}, V., {Sauvage}, J.-F., {Fusco}, T., and
  {Farinato}, J., ``{Laboratory demonstration of focal plane wavefront sensing
  using phase diversity: a way to tackle the problem of NCPA in SHARK-NIR},''
  in [{\em Society of Photo-Optical Instrumentation Engineers (SPIE) Conference
  Series}{\nolinebreak\hspace{0.1em}]},  {\em Society of Photo-Optical
  Instrumentation Engineers (SPIE) Conference Series} {\bf 11447},  1144753
  (Dec. 2020).

\bibitem{2020SPIE11447E..50M}
{Marafatto}, L., {Biondi}, F., {Carolo}, E., {Umbriaco}, G., {Bergomi}, M., {De
  Pascale}, M., {Greggio}, D., {Lessio}, L., {Radhakrishnan Santhakumari},
  K.~K., {Vassallo}, D., {Viotto}, V., and {Farinato}, J., ``{SHARK-NIR:
  challenges and solutions of a high contrast imager alignment},'' in [{\em
  Society of Photo-Optical Instrumentation Engineers (SPIE) Conference
  Series}{\nolinebreak\hspace{0.1em}]},  {\em Society of Photo-Optical
  Instrumentation Engineers (SPIE) Conference Series} {\bf 11447},  1144750
  (Dec. 2020).

\bibitem{2020SPIE11447E..4RU}
{Umbriaco}, G., {Carolo}, E., {Vassallo}, D., {Greggio}, D., {Marafatto}, L.,
  {Farinato}, J., {Baudoz}, P., {Bergomi}, M., {Biondi}, F., {Lessio}, L.,
  {Carlotti}, A., {Ragazzoni}, R., and {Viotto}, V., ``{The optical alignment
  of the coronagraphic masks of SHARK-NIR: paving the way for exoplanets
  detection and characterization},'' in [{\em Society of Photo-Optical
  Instrumentation Engineers (SPIE) Conference
  Series}{\nolinebreak\hspace{0.1em}]},  {\em Society of Photo-Optical
  Instrumentation Engineers (SPIE) Conference Series} {\bf 11447},  114474R
  (Dec. 2020).

\bibitem{2018SPIE10705E..16V}
{Vassallo}, D., {Farinato}, J., {Sauvage}, J.~F., {Fusco}, T., {Greggio}, D.,
  {Carolo}, E., {Viotto}, V., {Bergomi}, M., {Marafatto}, L., {Baruffolo}, A.,
  and {De Pascale}, M., ``{Validating the phase diversity approach for sensing
  NCPA in SHARK-NIR, the second-generation high-contrast imager for the Large
  Binocular Telescope},'' in [{\em Modeling, Systems Engineering, and Project
  Management for Astronomy VIII}{\nolinebreak\hspace{0.1em}]},  {Angeli}, G.~Z.
  and {Dierickx}, P., eds., {\em Society of Photo-Optical Instrumentation
  Engineers (SPIE) Conference Series} {\bf 10705},  1070516 (July 2018).

\bibitem{oarpaf-ricci}
{Ricci}, D. et~al., ``{Toward the remotization and robotization of the OARPAF
  telescope},'' in [{\em {Observatory Operations: Strategies, Processes, and
  Systems IX}}{\nolinebreak\hspace{0.1em}]},  {\em \procspie} {\bf 12186-24}
  (2022).

\end{thebibliography}

\end{document}